# Modeling of Multimodal Scattering by Conducting Bodies in Quantum Optics: the Method of Characteristic Modes

Gregory Ya. Slepyan[(1)], Dmitri Mogilevtsev[(2)], Ilay Levie[(1)], and Amir Boag[(1)]

*(1) School of Electrical Engineering, Tel Aviv University, Tel Aviv 69978, Israel*
*(2) Institute of Physics, Belarus National Academy of Sciences, Nezavisimosti Ave. 68, Minsk 220072, Belarus*

We propose a numerical technique for modeling the quantum multimode light scattering by a perfectly conducting body. Using the novel quantization technique, we give the quantum adaptation of the characteristic mode approach widely used in the classical electrodynamics. The method is universal with respect to the body's configuration, as well as its dimensions relative to the wavelength. Using this method and calculating the first- and the second-order field correlation functions, we demonstrate how scattering affects quantum-statistical features of the field. As an example, we consider scattering of the two single-photon incident Gaussian beams on the cylinder with circular cross-section. We show that the scattering is accompanied by the two-photon interference and demonstrates the Hong-Ou-Mandel effect. It is shown, that the scattered two-photon field and its correlations are able to manifest a varying directive propagation, which is controllable by various means (angles of incidence, configuration of the body, relations between its sizes with the frequency). We expect that this method will be useful for designing quantum-optical devices.

## I.    INTRODUCTION

The recent progress in quantum information [1,2], quantum computing [3], and various types of quantum technologies [4] make it timely to develop theoretical and numerical methods for the analysis of realistic quantum optical elements and systems, such as quantum networks [5-7], quantum transmission lines [8], quantum antennas [9-14], and quantum sensing devices [15]. Early fundamentals-oriented research in quantum optics necessarily concentrated on simple optical instruments, such as semitransparent mirrors [16-18], beam splitters [16-18], and different types of interferometers [19]. As a result, a certain schematic paradigm was born, which became the framework for considering/designing different types of quantum devices. In many studies, the devices are based on the reflection-transmission of light comprising single (or very few) modal components (actually, two-port and four-port devices have been often considered as passive components). As a basic interference component, a single-mode beam splitter was widely employed and described by a series of models [3, 18, 20]. The maximally complete physical model [20] accounted for the non-perfect properties of the reflectors dictated by the losses (gain). Losses produce noise, which should be incorporated in physically motivated field quantization satisfying correct commutation relations for the electromagnetic (EM)-field operators.

In recent decades, the principles of classical antennas [21] and microwave engineering [22] were transferred to the visible light and stimulated development of nano-antennas [23-26]. Then, these concepts were combined with the quantum aspects of light, creating the quantum antennas [9-14]. On the other hand, the quantum-optical concepts were extended to rather low frequencies, looking promising for new applications of microwave photons in quantum informatics, quantum identification of the targets (in particular, quantum radars) [27, 28], and metrology [29]. In this process, the basic elements of these devices became multi-modal involving wave scattering and diffraction. In [30], the classical and quantum scattering cross sections are expectedly shown to be exactly equal in the linear regime (when the multi-photon scattering processes are negligible). However, this result by no means precludes the future development of the quantum theory of scattering. The reason is that quantum scattering is not confined by the variance of the mode distribution. It is characterized by the new degrees of freedom opened up in the possibility of controlling the quantum statistics of light, which have no analogs in the classical scattering. It leads to specifically quantum effects stemming from quantum correlations. To name a few, one can mention the directivity of single photons spontaneously emitted by a dense atomic cloud [31], directivity of photon pair emission in quantum antennas, and suppression of single-photon components by two-photon interference referred as Hong-Ou-Mandel effect [32-36]. One of the most important quantum aspects is entanglement of quantum states. It may be both the working principle in quantum computing and informatics [1, 2, 19] and unwanted parasitic factor governing the electromagnetic compatibility at the nanoscale [37]. As a representative example, one may note the generation of entanglement via diffraction by a screen with aperture [38, 39] (where its analogy with the beam splitter has been also deduced).

The quantum theory of diffraction in [38,39] corresponds to the intuitive approach of physical optics, whose classical analog is valid only for large apertures with sizes significantly larger than the wavelength (Huygens – Kirchhoff principle). Such approach of today's quantum optics appears to be similar to the early period of development of classical antennas and microwave engineering [21,22]: the broad-brush approximate approaches formed the intuition for the semi-empirical engineering design. On the other hand, an important part of theoretical framework was produced by the exact solutions via separation of variables in special coordinate systems for highly symmetric "canonical" scatterers [40, 41]. Also, advances in numerical algorithms opened the way to the robust computer modeling based on the solution of the boundary-value problems for Maxwell equations with arbitrary degree of accuracy [42].

One can prognosticate a similar tendency in the further development of quantum optics. Some recent papers [43,44] propose quantization techniques for exact solutions via separation of variables for spherically layered multimode systems [43] and parabolic mirrors [44]. Works [45,46] can be seen as the first attempts along the path of computer modeling via direct numerical techniques. There is no doubt that the rich set of analytic and numerical tools of the classical electrodynamics will be useful in this process. The main components of the numerical techniques in classical electrodynamics (method of moments/boundary element method for integral equations, finite differences, finite elements, etc.) is based on the so-called "discretization" of Maxwell equations. This term in general means i) the presentation of EM-field on the introduced basis; ii) obtaining the infinite-order matrix equation for the unknown weight coefficients; iii) truncating this matrix to the finite-size; and finally iv) numerical solution of the resulting system of linear equations. Therefore, the optimal choice of the basis highly important in the computational algorithm design.

One of the most efficient sets of basis functions in classical electrodynamics is generated in the method of characteristic modes [47-52]. Characteristic modes of EM-fields are defined as fields produced by the so called characteristic currents distributed on the surface of the scatterer. The characteristic currents correspond to the eigenmodes of surface integral operators, with the surface impedance in the capacity of a spectral parameter (instead of the resonant frequency in many types of conventional techniques). It may be considered as a special case of generalized method of eigenoscillations [53], in which some other values may be used as a spectral parameters too (for example, permittivity or permeability, one of the scatterer sizes, coefficient in radiation conditions at the infinity, etc.). It is important to note that the spectrum of characteristic modes is always discrete. In the characteristic mode basis, the scattering matrix has a diagonal form. Therefore, the description of scattering by a lossless body in the basis of characteristic modes leads only to the phase shifts, in general, different for each mode. For the canonical scatterers, the characteristic modes (CMs) can be founded analytically via separation of variables, while for general configurations the CMs can be obtained numerically. Thus, this method is a computationally efficient approach that is universal with respect to the scatterer configuration. This makes characteristic mode approach promising for generalization to quantum optics, as described in this work.

In contrast to the classical optics, the Maxwell equations for the quantum light are reformulated in terms of the operators for EM-field. The conventional type of such reformulation is based on the decomposition of the field into plane waves while exchanging the classical weight coefficients by the pairs of creation-annihilation operators [16,17,54,55]. While appealing for its simplicity, the plane wave basis is not always efficient for the physical analysis and numerical modeling of complicated multi-mode processes. Thus, a unitary transformation, yielding a new set of basis functions (and corresponding creation-annihilation operators) may be performed. As noted in [19], the optimal choice of the new modes is able to transform the quantum state, pure or mixed, to the form that is maximally efficient for considering and understanding its scattering properties. The modes of this basis were named the principal modes [19]. The principal modes in the meaning of [19] preserve the orthogonality and the canonical bosonic form of the commutation relations for the corresponding field operators. The basis of characteristic modes seems well suited for consideration of entanglement arising by scattering, since characteristic modes are physically distinguishable. The unique properties of characteristic modes mentioned above allow us consider them as a principal basis both from physical analysis and computational point of view.

In this paper we develop the method of characteristic modes for the scattering of quantum light. As a simple (but sufficiently physically rich) example of the method application, we consider a perfectly conducting scatterer in the form of an infinitely long perfectly conducting cylinder with arbitrary cross section. The impinging light is polarized along the axis of the cylinder.

The paper is organized as follows. In Section II, we give a short review of characteristic mode technique for classical electrodynamics. In Section III, we propose the canonical quantization of characteristic modes. In Section IV, we consider an example of application to the scattering of two Gaussian beams on a perfectly conducting cylinder, and discuss two-photon interference effects, in particular, manifestations of the Hong-Ou-Mandel effect for the pair of single-photon beams. We end the paper with some concluding remarks and an outlook in Section V.

## II. CHARACTERISTIC MODES FOR CLASSICAL FIELD: DEFINITIONS, NOTATIONS, PROPERTIES

For simplicity, we consider a two-dimensional (2D) multimode scattering problem of an E-polarized electromagnetic field by a perfectly conducting cylinder as shown in Fig. 1. Monochromatic incident field with $e^{-i\omega t}$ time dependence is assumed.

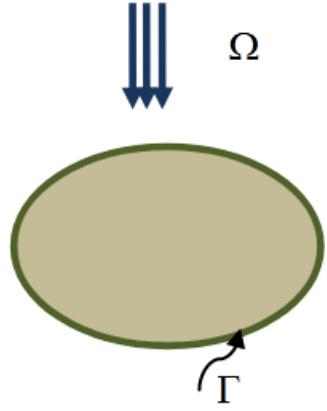

Figure 1. Scattering by a perfectly conducting cylinder, which is uniform and infinitely long (colored grey); The cross-section of the cylinder is of arbitrary shape. The area surrounding the scatterer and its boundary are denoted $\Omega$ and $\Gamma$, respectively.

The shape of the cylinder cross section is arbitrary (and its size is arbitrary compared to the wavelength of the incident field). The cylinder is infinite and uniform along the $z$-axis. In this case, the EM-field is completely defined by the longitudinal electric field component $E(\boldsymbol{\rho})$, which satisfiers the 2D Helmholtz equation

$$\nabla^2 E(\boldsymbol{\rho}) + k^2 E(\boldsymbol{\rho}) = 0, \quad \boldsymbol{\rho} \in \Omega \qquad (1)$$

where $k = \omega/c$ is the wavenumber, $c$ being the speed of light, accompanied by the boundary condition

$$E(\boldsymbol{\rho})\big|_{\Gamma} = 0 \qquad (2)$$

Here, $e^{-i\omega t}$ harmonic time dependence is assumed and suppressed.

Let us introduce the generalized impedance integral operator [47-52]

$$Z_k\{J\}(\mathbf{x}) = -i\omega\mu \oint_{\Gamma} G(\mathbf{x},\mathbf{x}';k) J(\mathbf{x}') d\Gamma' \qquad (3)$$

where $\mu$ is the permeability of the surrounding medium and $G(\mathbf{x},\mathbf{x}';k) = -\dfrac{i}{4} H_0^{(1)}(k|\mathbf{x}-\mathbf{x}'|)$ is the Green function of the 2D Helmholtz equation with $H_0^{(1)}(\cdot)$ being the Hankel function of the zeroth order and the first type and $\mathbf{x}, \mathbf{x}' \in \Gamma$. The characteristic currents $J_{nk}(\mathbf{x})$ are defined as the non-trivial solutions of equation

$$(R_k - iX_k)\{J_{nk}\} = (1 - i\lambda_n(k))R_k\{J_{nk}\} \tag{4}$$

where $R_k$ and $X_k$ are the real and imaginary parts of operator $Z_k\{J\}$ respectively, $\lambda_n = \lambda_n(k)$ is the eigenvalue of the *n*th characteristic mode. The eigenvalues' range is $-\infty < \lambda_n(k) < \infty$, with those of smaller absolute value being the more important for the description of the scattering process. Following [47-52], we order characteristic modes according to $|\lambda_1| \le |\lambda_2| \le ...|\lambda_n|... < \infty$ for a given value of $k$. The limit of eigenvalues $\lambda_n \to \infty$ for all *n* corresponds to the absence of scatterer. The characteristic currents are normalized such that

$$\oint_\Gamma J_{mk}(\mathbf{x})Z_k J_{nk}(\mathbf{x})d\Gamma = \delta_{mn}(1 - i\lambda_n(k)) \tag{5}$$

The *n*th field characteristic mode is defined as a field excited by the corresponding characteristic current and is given by

$$E_{nk}(\boldsymbol{\rho}) = -i\omega\mu \oint_\Gamma G(\boldsymbol{\rho},\mathbf{x}';k) J_{nk}(\mathbf{x}')d\Gamma' \quad \boldsymbol{\rho} \in \Omega \tag{6}$$

One of the most important properties of the characteristic modes is their orthogonality in the far zone (orthogonality of the radiation patterns) [47-52]. Following [47-52], we have for all modes at arbitrary frequencies

$$\oint_{\Gamma_\infty}(\mathbf{E}_{nk} \times \mathbf{H}^*_{mk}) \cdot \mathbf{n}d\Gamma = -\frac{1}{i\omega\mu}\oint_{\Gamma_\infty} E_{nk}\frac{\partial E^*_{mk}}{\partial \rho}d\Gamma = \delta_{mn} \tag{7}$$

where $\mathbf{H}_{mk}$ is the magnetic field of *m*th characteristic mode, $\mathbf{n}$ is the unit normal vector with respect to the contour $\Gamma$. A characteristic mode in the far field zone can be expressed as

$$E_{nk}(\boldsymbol{\rho}) \underset{k\rho \to \infty}{\sim} \frac{C_{nk}}{\pi}\sqrt{\frac{1}{ik\rho}}i^{-n}e^{ik\rho}\Phi_{nk}(\varphi) \tag{8}$$

where $\Phi_{nk}(\varphi)$ is the complex radiation pattern and $C_{nk}$ is the normalization constant. The radiation pattern is frequency dependent, while we often omit this dependence for conciseness. We will normalize the radiation patterns according to

$$\frac{1}{2\pi}\int_0^{2\pi}\Phi_{nk}(\varphi) \cdot \Phi^*_{mk}(\varphi)d\varphi = \delta_{mn} \tag{9}$$

Using (8), we obtain from (7) for the normalization constant $C_{nk} = C_k = \sqrt{\dfrac{\pi\omega\mu}{2}}$.

## III. QUANTIZATION OF CHARACTERISTIC MODES

In this section, we will consider the quantization of the fields in the far zone using asymptotic relation (8). Let us introduce the basis functions of characteristic modes

$$f_{nk}(\mathbf{\rho}) = \left[\frac{E_{nk}(\mathbf{\rho}) + E_{nk}^*(\mathbf{\rho})}{2} + P_{nn}(k)E_{nk}(\mathbf{\rho})\right] = \frac{1}{2}\left(S_{nn}(k)E_{nk}(\mathbf{\rho}) + E_{nk}^*(\mathbf{\rho})\right) \quad (10)$$

where the quantities $P_{nn}(k)$ and $S_{nn}(k)$ are the diagonal elements of the so-called perturbation and scattering matrices, respectively [47-52]. These matrices are expressed in the terms of eigenvalues $\lambda_n(k)$ as

$$\underline{P}(k) = \begin{pmatrix} \cdots & & 0 \\ & -\dfrac{1}{1-i\lambda_n(k)} & \\ 0 & & \cdots \end{pmatrix} \quad (11)$$

$$\underline{S}(k) = \begin{pmatrix} \cdots & & 0 \\ & -\dfrac{1+i\lambda_n(k)}{1-i\lambda_n(k)} & \\ 0 & & \cdots \end{pmatrix} \quad (12)$$

The limit of the matrices in the case of all $\lambda_n \to \infty$ is $\underline{P} \to 0, \underline{S} \to \underline{1}$. The basis functions (9) are characterized by two indices, one of which is continuous (the wavenumber $k = \omega_k/c$) while the other - discrete (the number of characteristic mode at a given frequency). Perturbation matrix (11) directly characterizes the scattering via the field perturbation by the cylinder (note that this perturbation is not assumed to be small), while the basis functions (10) satisfy the boundary condition over the whole surface of the scatterer. The scattering matrix $\underline{S}$ is unitary. Both matrices (11) and (12) are diagonal in the basis of characteristic modes. The lossless scattering process in the language of characteristic modes adds up to the phase shift, dependent on the mode number. As the mode index $n$ increases, $\lambda_n \to \infty$, which indicates decreasing perturbation of the given characteristic mode by the scattering process [47-52].

As usual, the general field operators are decomposed into the positive- and negative-frequency components according to $\hat{E}(\mathbf{\rho}) = \hat{E}^+(\mathbf{\rho}) + \hat{E}^-(\mathbf{\rho})$ (and the same for the vector potential $\hat{A}(\mathbf{\rho})$) [16,17]. The corresponding operators are

$$\hat{A}^+(\mathbf{\rho},t) = \int_0^\infty \frac{\mathrm{E}_k}{\omega_k} e^{-i\omega_k t} \sum_n \hat{c}_{nk} f_{nk}(\mathbf{\rho}) dk, \quad (13)$$

$$\hat{E}^-(\mathbf{\rho},t) = -i\int_0^\infty \mathrm{E}_k^* e^{i\omega_k t} \sum_n \hat{c}_{nk}^+ f_{nk}^*(\mathbf{\rho}) dk, \quad (14)$$

with commutation relations for creation-annihilation operators $\left[\hat{c}_{nk}, \hat{c}_{n'k'}^+\right] = \delta_{nn'}\delta(k-k')$ (all other pairs of operators commute), and $\mathrm{E}_k = \sqrt{\hbar\omega_k c/\pi}$ is the normalization coefficient.

Let us prove that operators (13), and (14) satisfy the canonical equal-time commutation relation

$$\left[\hat{A}^+(\boldsymbol{\rho},t),\hat{E}^-(\boldsymbol{\rho}',t)\right]=-\frac{i}{2\varepsilon}\hbar\delta(\boldsymbol{\rho}-\boldsymbol{\rho}') \qquad (15)$$

Using the commutation relation for the creation-annihilation operators, we obtain

$$\left[\hat{A}^+(\boldsymbol{\rho},t),\hat{E}^-(\boldsymbol{\rho}',t)\right]=-\frac{i\hbar}{\pi}\sum_n\int_0^\infty f_{nk}(\boldsymbol{\rho})f_{nk}^*(\boldsymbol{\rho}')dk \qquad (16)$$

Employing relation (10) for the basis functions, asymptotic representation (8), and well-known properties $S_{nn}(-k)=S_{nn}^*(k)$ and $\Phi_{n,-k}(\varphi)=\Phi_{nk}^*(\varphi)$ [56], we can rewrite (16) in the following form:

$$\left[\hat{A}^+(\boldsymbol{\rho},t),\hat{E}^-(\boldsymbol{\rho}',t)\right]=-\frac{i\hbar}{8\pi^2\varepsilon}\frac{1}{\sqrt{\rho\rho'}}\times$$

$$\sum_n\int_0^\infty[e^{ik(\rho-\rho')}\Phi_{nk}(\varphi)\Phi_{nk}^*(\varphi')+i^{-(2n+1)}S_{nn}(k)e^{ik(\rho+\rho')}\Phi_{nk}(\varphi)\Phi_{nk}(\varphi')]dk+\text{c.c.}= \qquad (17)$$

$$-\frac{i\hbar}{8\pi^2\varepsilon}\frac{1}{\sqrt{\rho\rho'}}\sum_n\int_{-\infty}^\infty[e^{ik(\rho-\rho')}\Phi_{nk}(\varphi)\Phi_{nk}^*(\varphi')+i^{-(2n+1)}S_{nn}(k)e^{ik(\rho+\rho')}\Phi_{nk}(\varphi)\Phi_{nk}(\varphi')]dk$$

In the second term of (17), we can define a quantity

$$\Theta(\varphi,\varphi';k)=\sum_n(-1)^n S_{nn}(k)\Phi_{nk}(\varphi)\Phi_{nk}(\varphi')$$

which belongs to the class of general susceptibilities [56]. It may be analytically continued to the complex $k$-plane and possesses the corresponding analytical properties [56]. As follows from the causality principle, this function is analytic in the upper half of the $k$-plane. Thus, the integration contour over the real axis in the second term of (17) may be closed by a semi-circle of infinite radius in upper half-plane. Using Cauchy theorem, we can see that the second term in the integrand in (17) does not contribute to the value of the commutator. Using the well-known identities

$$(2\pi)^{-1}\sum_n\Phi_{nk}(\varphi)\Phi_{nk}^*(\varphi')=\delta(\varphi-\varphi') \text{ and } (2\pi)^{-1}\int_{-\infty}^\infty e^{ik(\rho-\rho')}dk=\delta(\rho-\rho'),$$

we obtain $\left[\hat{A}^+(\boldsymbol{\rho},t),\hat{E}^-(\boldsymbol{\rho}',t)\right]=-i\hbar\rho^{-1}\delta(\rho-\rho')\delta(\varphi-\varphi')/2\varepsilon$, which is exactly the correct commutation relation (15).

In the case of a monochromatic field, we account for all characteristic modes of the same frequency. In this case, we obtain from (13) and (14) the following expressions for the operators of field amplitudes

$$\hat{E}^{-}(\boldsymbol{\rho}) = -i\sqrt{\frac{\hbar\omega}{\pi}} \sum_{n} \hat{c}_{n}^{+} f_{nk}^{*}(\boldsymbol{\rho})$$
$$\hat{E}^{+}(\boldsymbol{\rho}) = i\sqrt{\frac{\hbar\omega}{\pi}} \sum_{n} \hat{c}_{n} f_{nk}(\boldsymbol{\rho}) \tag{18}$$

The operators $\hat{c}_n, \hat{c}_n^+$ satisfy the commutation relations $\left[\hat{c}_n, \hat{c}_{n'}^+\right] = \delta_{nn'}$ (while all other pairs commute).

Similar relations may be written separately for incoming and outgoing components of the field operators:

$$\hat{E}^{+}_{\text{incom}}(\boldsymbol{\rho}) = i\sqrt{\frac{\hbar\omega}{\pi}} \sum_{n} \hat{c}_{n} E_{n}^{*}(\boldsymbol{\rho})$$
$$\hat{E}^{+}_{\text{outgo}}(\boldsymbol{\rho}) = i\sqrt{\frac{\hbar\omega}{\pi}} \sum_{n'} \hat{c}_{n} S_{nn}(k) E_{n}(\boldsymbol{\rho}) \tag{19}$$

The electric field in the far field zone may be represented in terms of radiation patterns using relation (8).

This section, actually, gives a complete general recipe for describing scattering of an incident quantum state by an arbitrary shaped lossless scatterer described by the set of its characteristic modes. Indeed, the scattering problem reduces to the representation of the incident field in terms of the characteristic modes. Formally, if the positive-frequency part of the incident field is represented in the form (19) using another set of basis functions with orthonormal radiation patterns $\Upsilon_n(\varphi)$ and the set of annihilation operators $\hat{a}_m$, one can obtain

$$\hat{a}_m = \sum_n V_{mn} \hat{c}_n \tag{20}$$

where the transformation coefficients $V_{mn} = \frac{1}{2\pi} \int_{S_\infty} \Phi_n(\varphi) \Upsilon_m^*(\varphi) d\varphi$ are defined through the overlap of the radiation patterns from both sets. The unitary transformation described by Eq. (20) allows one to express an arbitrary state in terms of the characteristic modes.

## IV. SCATTERING OF TWO BEAMS BY A CYLINDER

### A. Double-mode single-photon field

First, we would like to note that despite the seeming simplicity of representation (20), using it for an arbitrary configuration of the incident field might be quite cumbersome. In addition, considerable complications might arise in the calculation of the resulting characteristic mode state for a large number of photons in the incident field. Notice that the state transformation akin to the one described by Eq. (20), boson sampling leads to the non-polynomial complexity problem [57].

In this section, we implement a methodologically not so simple, but more straightforward and physical approach for the scattering of the two Gaussian beams by a cylindrical scatterer with a circular cross section. Due to the additional rotational symmetry, functions given by relation (10) become orthogonal, and the products of characteristic modes with different indices do not contribute to the commutator (15). Thus, the correct commutation relation (15) is satisfied for the fields (13), (14) over the whole space, and not only in the far zone.

Our analysis is based on the concept of principal modes suggested in [19] and a representation of the principal modes using the characteristic mode basis. The principal modes are defined as a special basis in which the first order coherency matrix $\mathbf{\Gamma}^{(1)}$ consists of a square $p \times p$ non-zero diagonal block surrounded by zeros [19]. Vice versa, if a matrix $\mathbf{\Gamma}^{(1)}$ has the form defined above, the number of photons in $n$th mode with $n > p$ is zero, which implies that this mode appears in the vacuum state.

Let us begin from a single-photon state of a single mode. As has been shown in [19], all single-photon states are actually single mode states. Let us introduce the state, which is the coherent superposition of single photon states in characteristic modes with the probability amplitudes equal to the excitation coefficients. It reads

$$|1:\mathbf{v}_1\rangle = \sum_n V_n^* |0_1, 0_2, ..., 1_n, ...\rangle \tag{21}$$

where $V_n$ are the excitation coefficients defined in Eq. (20) and satisfying the normalization condition $\sum_n |V_n|^2 = 1$. This state may be associated with the action of operator $\hat{a}^+ = \sum_n V_n^* \hat{c}_n^+$ on vacuum, i.e., $|1:\mathbf{v}_1\rangle = \hat{a}^+ |0\rangle$. As follows from the commutation relations for operators $\hat{c}_n, \hat{c}_n^+$ and the normalization of the excitation coefficients, this is a bosonic operator which satisfies the commutation relation $\left[\hat{a}, \hat{a}^+\right] = 1$. It follows, that (21) is indeed a single photon state.

Let us find the EM-field of the mode carrying the single photon. We introduce two partial modes given by linear combinations of the characteristic modes

$$\mathbf{v}_{1,\text{incom}}(\mathbf{\rho}) = \sqrt{\frac{1}{2\pi\omega\mu}} \sum_n V_n^* E_n^*(\mathbf{\rho}) \tag{22}$$

$$\mathbf{v}_{1,\text{outgo}}(\mathbf{\rho}) = \sqrt{\frac{1}{2\pi\omega\mu}} \sum_n V_n^* S_{nn} E_n(\mathbf{\rho}) \tag{23}$$

and the total field $\mathbf{v}_1 = \mathbf{v}_{1,\text{incom}} + \mathbf{v}_{1,\text{outgo}}$ given by a sum of relations (22) and (23). This field may be considered as a first element of a new modal basis $\{\mathbf{v}_n\}$, which is completed by the orthogonal states. The wavefunction in this basis can be expressed in the form of a tensor product $|\mathbf{\Psi}\rangle = |1:\mathbf{v}_1\rangle \otimes ... \otimes |0:\mathbf{v}_n\rangle \otimes ...$. It means, that this wavefunction may be indeed associated with the principal mode, whose creation operator is $\hat{a}^+$. The fields $\mathbf{v}_{1,\text{incom}}$, $\mathbf{v}_{1,\text{outgo}}$ correspond to the incoming and outgoing components of the total field, respectively. Therefore, the single principal mode exactly describes the scattering of the single photon.

Now let us introduce the principal modes for the superposition of two single-photon states. We are interested in the situation schematically depicted in Fig. 2: two spatially separate beams denoted $e_1(\boldsymbol{\rho})$ and $e_2(\boldsymbol{\rho})$ (for example, two Gaussian beams) are impinging on the cylindrical scatterer. These beams may be presented in terms of the functions (22), (23) as $e_j(\boldsymbol{\rho}) = \mathbf{v}_{j,\text{incom}} + \mathbf{v}_{j,\text{outgo}}$ with $S_{nn} = 1$.

Generally, these two modes are not orthogonal. So, if one tries to introduce a single-photon states of each beam in the manner it was done above for the single-photon field, i.e., to take

$$\hat{a}_j^+ = \sum_n V_{nj}^* \hat{c}_n^+ \qquad j = 1, 2 \tag{24}$$

with the coefficients $V_{nj}$ describing representation of the each mode in terms of the characteristic modes while satisfying the normalization condition

$$\sum_n |V_{n1}|^2 = \sum_n |V_{n2}|^2 = 1 \tag{25}$$

and assumes the state as

$$|1_1, 1_2\rangle = \hat{a}_1^+ \hat{a}_2^+ |0\rangle, \tag{26}$$

one would observe that the photon creation operators do not correspond to independent modes and satisfy the following commutation relations

$$\left[\hat{a}_i, \hat{a}_j^+\right] = \delta_{ij} + (1 - \delta_{ij}) \sum_n V_{nj}^* V_{ni} \tag{27}$$

To simplify the discussion, we assume that the first photon is in mode 1, but the second one is actually in the spatial mode orthogonal to the first one, i.e., if we introduce the second mode with the help, for example, of the orthogonalization relation

$$\tilde{e}_2(\boldsymbol{\rho}) \propto e_2(\boldsymbol{\rho}) - \mu_{12} e_1(\boldsymbol{\rho}), \tag{28}$$

where the overlap between two modes is

$$\mu_{12} = \int e_2(\boldsymbol{\rho}) e_1^*(\boldsymbol{\rho}) d\Omega, \tag{29}$$

with integration over the whole space. Then the creation operators $\hat{a}_1^+, \hat{a}_2^+$ would create photons in the orthogonal modes, and the state (26) would truly correspond to just one photon in each independent mode impinging on the scatterer. We adopt this situation for the illustration presented below. Notice that when the modes are well spatially separated (as illustrated in Fig. 2), one can hardly expect a significant overlap between modes.

### B. Scattering of two beams by the cylinder with a circular cross-section

In this subsection, we will consider the scattering of two single-photon beams by a circular cylinder based on the formalism developed above (the problem configuration is shown in Fig. 2). For every beam, we use a Gaussian beam model with the width large compared to the wavelength. The incident field of the beam in the local Cartesian coordinate system in the waist plane $x = x_0$ is given by $e_{inc}(-x_0, y) = E_0 e^{-\beta^2 y^2}$. The parameter $\beta = 1/\sqrt{2W}$ where $W$ is the width of the beam in the plane $x = x_0$. We assume a wide beam illumination, i.e. $W \gg \lambda$. Let us also assume that the two beams are identical, except for the second beam being rotated by angle $\theta$ with respect to the first one (see Fig. 2).

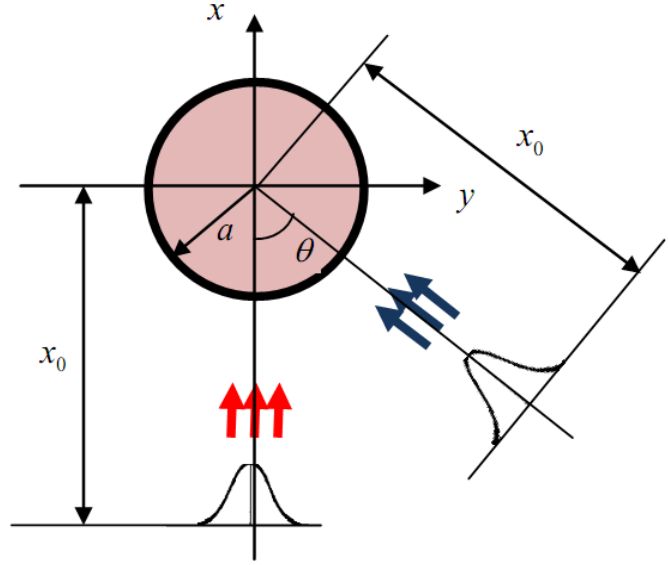

Fig. 2. Scattering of the two beams by a cylinder with a circular cross-section.

To characterize the scattering process, we will calculate the first order correlation function

$$G^{(1)}(\mathbf{\rho}_1, \mathbf{\rho}_2) = \langle \hat{E}^-(\mathbf{\rho}_1) \hat{E}^+(\mathbf{\rho}_2) \rangle \quad (30)$$

and the second-order normalized correlation function

$$g^{(2)}(\mathbf{\rho}_1, \mathbf{\rho}_2) = \frac{\langle \hat{E}^-(\mathbf{\rho}_1) \hat{E}^-(\mathbf{\rho}_2) \hat{E}^+(\mathbf{\rho}_2) \hat{E}^+(\mathbf{\rho}_1) \rangle}{\langle \hat{E}^-(\mathbf{\rho}_1) \hat{E}^+(\mathbf{\rho}_1) \rangle \langle \hat{E}^-(\mathbf{\rho}_1) \hat{E}^+(\mathbf{\rho}_1) \rangle} \quad (31)$$

(the angular brackets mean the expectation value with respect to the quantum state of the incident field). The field in Eqs. (30), (31) is given by $\hat{E}^-(\mathbf{\rho}) = -i\omega \sqrt{\frac{\hbar \mu}{2}} \left( \hat{a}_1^+ \mathbf{v}_1^*(\mathbf{\rho}) + \hat{a}_2^+ \tilde{\mathbf{v}}_2^*(\mathbf{\rho}) \right)$ where $\tilde{\mathbf{v}}_2^*(\mathbf{\rho})$ is the orthogonal to $\mathbf{v}_1^*(\mathbf{\rho})$ in accordance with Eqs. (28) and (29).

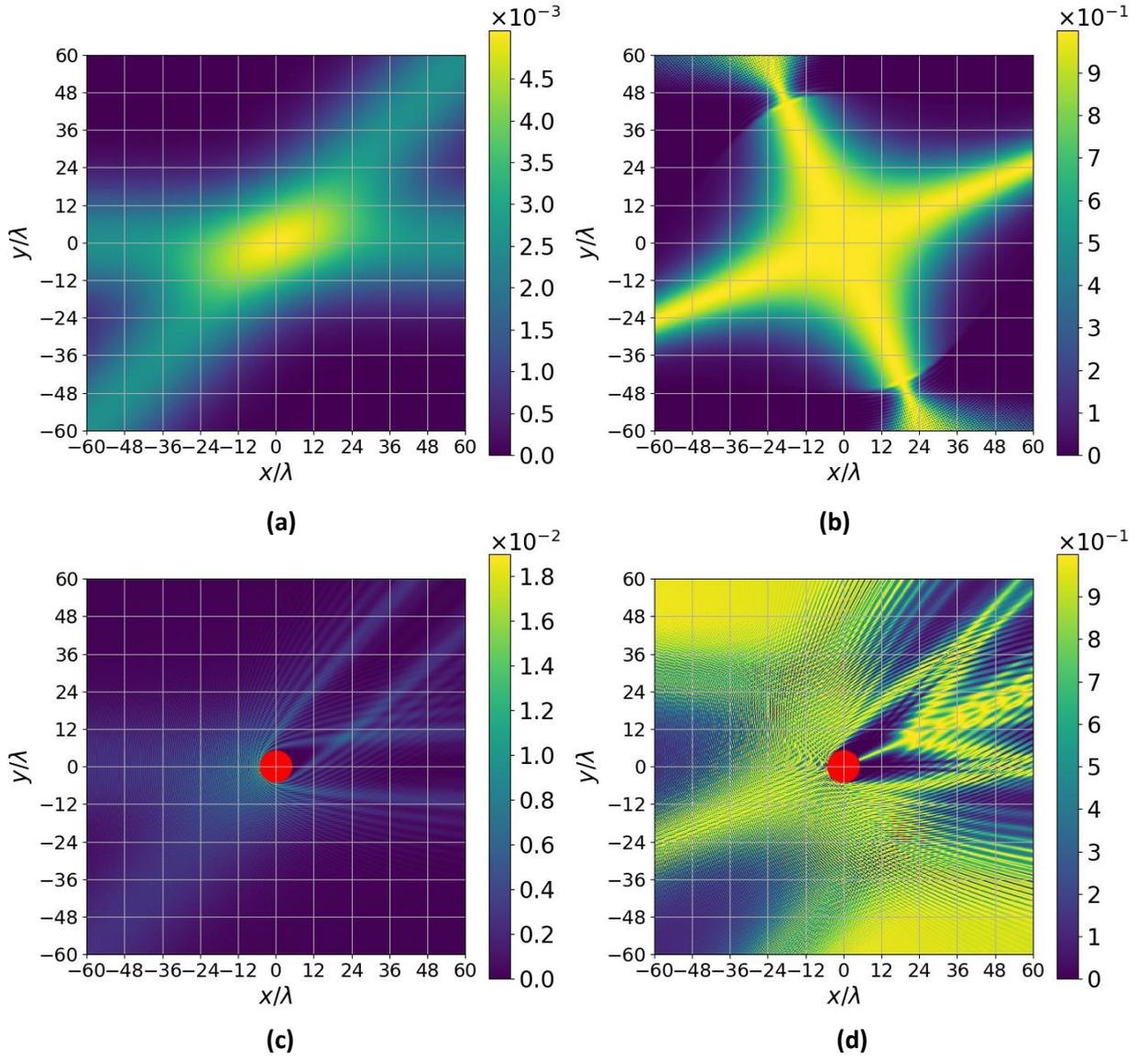

Fig. 3. Scattering of two single-photon Gaussian beams: (a) and (c) intensity $G^{(1)}(\boldsymbol{\rho}_1,\boldsymbol{\rho}_2)\big|_{\boldsymbol{\rho}_1=\boldsymbol{\rho}_2=\boldsymbol{\rho}}$ in the absence and in presence of the scatterer, respectively; (b) and (d) equal-point normalized second-order correlation function $g^{(2)}(\boldsymbol{\rho}_1,\boldsymbol{\rho}_2)\big|_{\boldsymbol{\rho}_1=\boldsymbol{\rho}_2=\boldsymbol{\rho}}$ in the absence and in presence of the scatterer, respectively. The cylinder colored red.

Toward using the formalism developed in the previous subsection, we must; i) introduce the doublet of the principal modes; ii) calculate the characteristic modes; iii) calculate the set of excitation coefficients. The characteristic modes for the circular cross section may be calculated analytically using the separation of variables [52]. The first of the principal modes is identified with one of Gaussian beams, while the second is obtained via orthogonalization of the Gaussian beams. The excitation coefficients for the Gaussian beam in the basis of cylindrical modes are given in Appendix.

In our example, the two single-photon Gaussian beams impinge from the left and lower left sides as illustrated in Fig. 3. Here, we choose the following parameter values: $a = 5\lambda$, $\theta = 45°$, $x_0 = 0$, $\beta = 1/25\lambda$. In the first example, the beams are independent, i.e., the wave-function of incident field is given by $|\Psi\rangle_{in} = |1:\mathbf{v}_1\rangle \otimes |1:\tilde{\mathbf{v}}_2\rangle$. Field intensity $G^{(1)}(\mathbf{\rho}_1,\mathbf{\rho}_2)\big|_{\mathbf{\rho}_1=\mathbf{\rho}_2=\mathbf{\rho}}$ and the equal-point normalized second-order correlation function $g^{(2)}(\mathbf{\rho}_1,\mathbf{\rho}_2)\big|_{\mathbf{\rho}_1=\mathbf{\rho}_2=\mathbf{\rho}}$ in the presence of the scatterer are shown in Figs. 3(c) and 3(d), respectively. The free-space interference of two Gaussian beams in the same quantum state is also depicted in Figs. 3(a,b). As one can see, the scattering dramatically changes the quantum statistical properties of light. Considering the intensity distribution of the fields in Fig. 3(c) and the normalized second-order correlation function Fig. 3(d), one can make a couple of interesting observations. First: there are single-photon shadows behind the cylinder where the field from only one of beams is present. Second observation is the manifestation of the Hong-Ou-Mandel effect. Arrow-shaped region behind the cylinder is the one where bright color corresponds to a considerable average number of photons, and relatively high probability of two-photon registration. For two detectors placed in such "bright" regions, one would have high probability of simultaneous registration of photons.

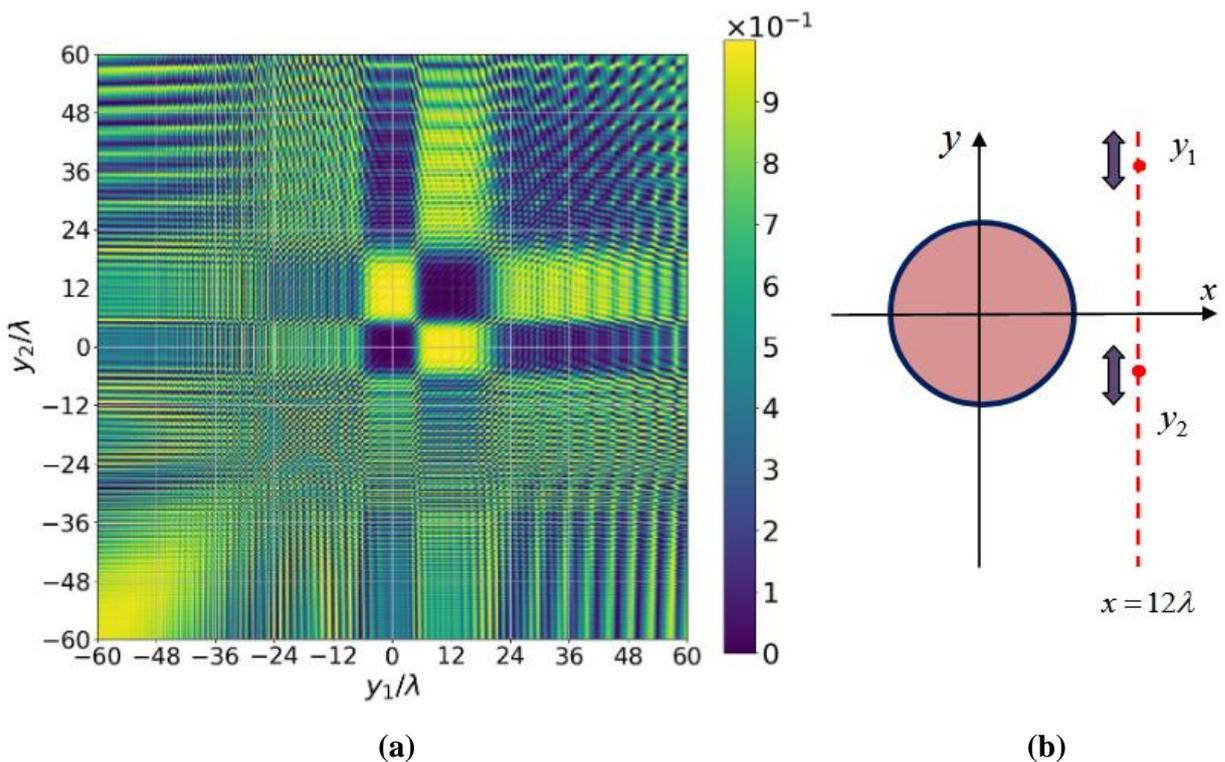

**(a)** **(b)**

Fig. 4. Scattering of two single-photon Gaussian beams: (a) non-equal-point normalized second-order correlation function $g^{(2)}(\mathbf{\rho}_1,\mathbf{\rho}_2)$; (b) location of the points $\mathbf{\rho}_j = (x_j; y_j), j = 1,2$ where $x_1 = x_2 = 12\lambda$, $\lambda$ being the wavelength, while $y_{1,2}$ are the variables along the axes in (a).

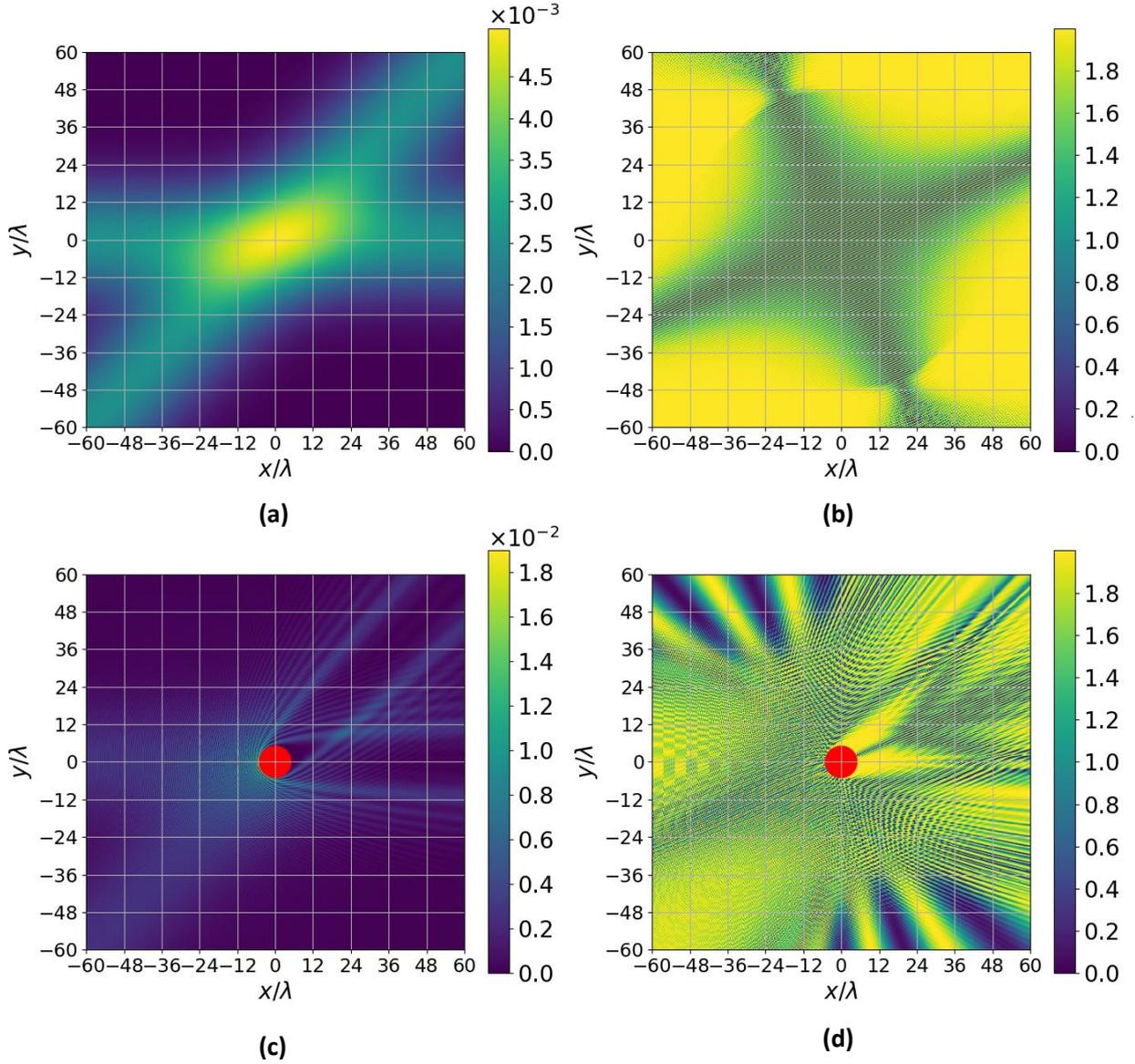

Fig. 5. Scattering of two Gaussian beams in the entangled state $|2002\rangle$: (a) and (c) intensity $G^{(1)}(\boldsymbol{\rho}_1,\boldsymbol{\rho}_2)\big|_{\boldsymbol{\rho}_1=\boldsymbol{\rho}_2=\boldsymbol{\rho}}$ in the absence and in presence of the scatterer, respectively; (b) and (d) equal-point normalized second-order correlation function $g^{(2)}(\boldsymbol{\rho}_1,\boldsymbol{\rho}_2)\big|_{\boldsymbol{\rho}_1=\boldsymbol{\rho}_2=\boldsymbol{\rho}}$ in the absence and in presence of the scatterer, respectively. The cylinder colored red.

It is important to note, that the qualitative behavior of the Hong-Ou-Mandel effect differs from its simplest manifestation in the $2\times 2$ beam-splitter [16-18]. The reason being the multimode origin of scattering field. As a result, priory two photons of the scattered field propagate jointly in the area of one of the principal modes (wave beams). However, these modes are strongly confined and have pronounced directive properties (for example, similar to the lobes of radiation patterns in antennas). The directions of joint photons propagation are effectively controllable via the configuration of scatterer, the angles of incidence, and the wavelength of the incident field. It opens the way for using the Hong-Ou-Mandel effect in quantum antennas with the complex spatial structure of the radiated field.

The case of the scattering of single-photon double-mode entangled state (the wave-function of incident field is $|\Psi\rangle_{in} = (|2:\mathbf{v}_1\rangle \otimes |0:\tilde{\mathbf{v}}_2\rangle + |0:\mathbf{v}_1\rangle \otimes |2:\tilde{\mathbf{v}}_2\rangle)/\sqrt{2}$) is shown in Fig. 5.

## V. CONCLUSION AND OUTLOOK

In this paper we developed a general numerical technique for modeling of the quantum light scattering. The method is based on the characteristic mode technique, widely used in the classical antennas theory and microwave engineering [52]. The method is universal with respect to the configuration of scatterer, structure of the incident field, and the scatterer dimensions relative to the wavelength. Toward computing the observable quantities (such as the correlation functions of different orders) one needs to calculate the characteristic modes for a given scatterer and the set of excitation coefficients. Such calculations may be completely done by means of classical electrodynamics. Thus, the developed method will be available in the near future for various quantum applications based on the computational tools of the classical electromagnetics.

In this paper, a relatively simple formulation of this method is demonstrated on the case of a the perfectly conducting 2D-scatterer of arbitrary cross section. The developed method may be rather directly modified for other types of the problems, for which the method of characteristic modes have been developed in the classical case (dielectric bodies, 3D-bodies, *N*-port circuits, coupling between volumes through an aperture).

As a simplest example of application, we consider the scattering of a pair of Gaussian quantum beams on a cylinder with a circular cross-section. Light intensity and the normalized second-order correlation function are calculated. Two quantum states of the incident light: i) a pair of single-photon independent beams; ii) the two beams in the double-photon entangled state are considered. It is shown, that scattering opens the way of control of quantum-statistical properties of light, such as producing, or suppressing the photonic entanglement.

A manifestation of Hong-Ou-Mandel effect in the scattering is demonstrated. It is shown that it qualitatively different as compared to its simplest manifestation in the $2\times 2$ beam-splitter [16-18] due to the pronounced directive properties. This opens the way for its use in quantum antennas.

## APPENDIX: APPROXIMATE EXCITATION COEFFICIENTS FOR GAUSSIAN BEAM

The purpose of this Appendix is the calculation of: i) characteristic modes for a circular cylinder and ii) the excitation coefficients for Gaussian beam. The operator of the Gaussian beam will be presented in the form of the single mode. The electric field operator for the principal mode is given by

$$\hat{E}^-(\boldsymbol{\rho}) \approx -i\omega \sqrt{\frac{\hbar\mu}{\pi}} \hat{a}^+ \sum_n V_n F_n^*(\boldsymbol{\rho}) \tag{A1}$$

where $\hat{a}^+$ is the creation operator of bosonic type. The characteristic functions for the circular cylinder scatterer are founded simply via the separation of variables in the cylindrical coordinate system. The final result reads

$$F_n(\mathbf{\rho}) = \frac{1}{\sqrt{2\pi}}\left(J_n(k\rho) + P_{nn}H_n^{(1)}(k\rho)\right)e^{-in\varphi} \tag{A2}$$

and $P_{nn} = -J_n(ka)/H_n^{(2)}(ka)$ is the diagonal element of perturbation matrix (see Eq. (11)) found from the boundary condition of the perfectly conducting surface of the cylinder. The functions (A2) are related to functions (10) as $f_n(\mathbf{\rho}) = \sqrt{\omega\mu}F_n(\mathbf{\rho})$.

The field in the far zone is a cylindrical wave

$$\hat{E}^+_{\text{outgo}} \xrightarrow{k\rho\to\infty} \frac{1}{\pi}\sqrt{\frac{1}{ik\rho}}e^{ik\rho}\hat{a}\Phi(\varphi) \tag{A3}$$

with the radiation pattern

$$\Phi(\varphi) = -i\omega\sqrt{\frac{\hbar\mu}{4\pi}}\sum_{n=-\infty}^{\infty}\frac{H_n^{(2)}(ka)}{H_n^{(1)}(ka)}i^{-n}V_n^* e^{-in\varphi} \tag{A4}$$

The next step of our calculations comprises the expression of coefficients $V_n$ in terms of the Gaussian beam parameters. Let us represent the Gaussian beam in the cylindrical coordinate system. In the Cartesian system, it is given by

$$e_{\text{inc}}(x,y) = \frac{1}{2\pi}\int_{-\infty}^{\infty} E(\alpha)e^{i\sqrt{k^2-\alpha^2}(x+x_0)}e^{i\alpha y}d\alpha \tag{A5}$$

where

$$E(\alpha) = \int_{-\infty}^{\infty} e_{\text{inc}}(-x_0, y)e^{-i\alpha y}d\alpha \tag{A6}$$

and $e_{\text{inc}}(-x_0, y) = E_0 e^{-\beta^2 y^2}$. The subscript "inc" means incident field which consists of both the incoming and outgoing fields of the same amplitudes. We assume that the inequality $|\beta\lambda| \ll 1$, which corresponds to the case of a wide beam is satisfied. Introducing cylindrical coordinates $x = \rho\cos(\varphi), y = \rho\sin(\varphi)$, $\alpha = k\sin(\mu)$, we rewrite (A1) as

$$e_{\text{inc}}(\rho,\varphi) = \frac{E_0}{2\beta\sqrt{\pi}}\int_{-\infty}^{\infty} e^{-\frac{\alpha^2}{4\beta^2}} e^{i\sqrt{k^2-\alpha^2}x_0} e^{ik\rho\cos(\varphi-\mu(\alpha))} d\alpha \tag{A7}$$

Using the well-known identity [58]

$$e^{ik\rho\cos(\varphi-\mu(\alpha))} = \sum_n \frac{1}{i^{-n}} e^{-in(\varphi-\mu(\alpha))} J_n(k\rho) \tag{A8}$$

we obtain

$$e_{inc}(\rho,\varphi) = \frac{E_0}{2\beta\sqrt{\pi}} \sum_n \frac{1}{i^{-n}} e^{-in(\varphi-\mu(\alpha))} J_n(k\rho) A_n \tag{A9}$$

with

$$A_n = \int_{-\infty}^{\infty} e^{-\frac{\alpha^2}{4\beta^2}} e^{i\sqrt{k^2-\alpha^2} x_0} e^{in\mu(\alpha)} d\alpha \tag{A10}$$

The relation (A10), in fact, is the incident field because it contains only the Bessel functions. For the wide beam case, the dominant contribution to the integration in (A10) is given by the region of small values of $\alpha$, for which we can use the approximation $\mu(\alpha) = \arcsin(\alpha/k) \approx \alpha/k$. As a result, we obtain: $A_n \approx (k\kappa/\beta)^{1/2} e^{ikx_0} e^{-\kappa n^2}$, $\kappa = \beta^2/k^2(1+i\xi)$, $\xi = 2\beta^2 x_0/k$.

For the excitation coefficient, we can take $V_n = Ce^{-\kappa n^2}/i^{-n}$ and find the normalization constant $C$ from the normalization condition $\sum_n |V_n|^2 = 1$ introduced above. Finally, we obtain

$$V_n \approx \frac{1}{\sqrt{1+2\sum_{m=1}^{\infty} e^{-2\operatorname{Re}(\kappa)m^2}}} \frac{1}{i^{-n}} e^{-\kappa n^2} \tag{A11}$$

To obtain a more accurate equation for excitation coefficients, one can use the approximation $\mu(\alpha) \approx \alpha/k + \frac{1}{6}(\alpha/k)^3 + \dots$. Then,

$$V_n \approx \frac{1}{\sqrt{1+2\sum_{m=1}^{\infty} |\Delta_m|^2 e^{-2\operatorname{Re}(\kappa)m^2}}} \frac{\Delta_n}{i^{-n}} e^{-\kappa n^2} \tag{A12}$$

where $\Delta_n \approx 1 - 2\kappa^2 n^2 - (4/3)\kappa^3 n^4$.

---

[33] T. Kobayashi, R. Ikuta, S. Yasui, S. Miki, T. Yamashita, H. Terai, T. Yamamoto, M. Koashi, and N. Imoto, Frequency-domain Hong–Ou–Mandel interference, Nat. Photon.|**10** (2016).

[34] E. Karimi, D. Giovannini, E. Bolduc, N. Bent, F. M. Miatto, M. J. Padgett, and R. W. Boyd, Exploring the quantum nature of the radial degree of freedom of a photon via Hong-Ou-Mandel interference, Phys. Rev. A **89**, 013829 (2014).

[35] Y. Zhang, S. Prabhakar, C. Rosales-Guzm´, F. S. Roux, E. Karimi, and A. Forbes, Hong-Ou-Mandel interference of entangled Hermite-Gauss modes, Phys. Rev. A **94**, 033855 (2016).

[36] M. Hiekkamäki, and R. Fickler, High-Dimensional Two-Photon Interference Effects in Spatial Modes, Phys. Rev. Lett, **126,** 123601 (2021).

[37] G. Slepyan, A. Boag, V. Mordachev, E. Sinkevich, S. Maksimenko, P. Kuzhir, G. Miano, M. E. Portnoi and A. Maffucci, Anomalous electromagnetic coupling via entanglement at the nanoscale, New J. Phys. **19**, 023014 (2017).

[38] Z. Xiao, R. N. Lanning, M. Zhang, I. Novikova, E. E. Mikhailov, and J. P. Dowling, Why a hole is like a beam splitter: A general diffraction theory for multimode quantum states of light, Phys. Rev. A **96** 023829 (2017).

[39] A. Z. Goldberg, and D. F. V. James, Entanglement generation via diffraction, Phys. Rev. A **100**, 042332 (2019).

[40] L. B. Felsen, and N. Marcuvitz, *Radiation and scattering of waves* (IEEE Press series on electromagnetic waves, Originally published: Englewood Cliffs, N.J.: Prentice-Hall, 1972).

[41] J. Jackson, *Classical Electrodynamics* (John Wiley & Sons, Inc., New York, 1962).

[42] A. F. Peterson, S. L. Ray, R. Mittra, *Computational Methods for Electromagnetics* (IEEE Press Series on Electromagnetic Wave Theory, Wiley-IEEE Press, 1997).

[43] D. Dzsotjan, B. Rousseaux, H. R. Jauslin, G. Colas des Francs, C. Couteau, and S. Guerin, Mode-selective quantization and multimodal effective models for spherically layered systems, Phys. Rev. A **94**, 023818 (2016).

[44] R. Gutiérrez-Jáuregui, and R. Jáuregui, Photons in the presence of parabolic mirrors, Phys. Rev. A **98**, 043808 (2018).

[45] D.-Y. Na, J. Zhu, W. C. Chew, and F. L. Teixeira, Quantum information preserving computational electromagnetic, Phys. Rev. A **102**, 013711 (2020).

[46] S. Savasta, O. Di Stefano, and R. Girlanda, Light quantization for arbitrary scattering systems, Phys. Rev. **65**, 043801 (2002).

[47] R. J. Garbacz, Modal expansions for resonance scattering phenomena, Proc. IEEE, **53**, 856 (1965).

[48] R. J. Garbacz, and R. H. Turpin, A generalized expansion for radiated and scattered fields, IEEE Trans. Antennas Propag., AP-**19**, 348 (1971).

[49] R. J. Garbacz, and D. M. Pozar, Antenna shape synthesis using characteristic modes, IEEE Trans. Antennas Propag., AP-**30,** 340 (1982).

[50] R. F. Harrington, and J. R. Mautz, Theory of characteristic modes for conducting bodies, IEEE Trans. Antennas Propag., AP-**19**, 622 (1971).

[51] R. F. Harrington, and J. R. Mautz, Computation of characteristic modes for conducting bodies, IEEE Trans. Antennas Propag., AP-**19**, 629 (1971).

[52] Y. Chen, and C.-F. Wang, *Characteristic Modes: Theory and Applications in Antenna Engineering*, (John Wiley & Sons, Inc. Published by John Wiley & Sons, Inc. 2015).

[53] M. S. Agranovich, B. Z. Katsenelenbaum, A. N. Sivov, and N. N. Voitovich, *Generalized Method of Eigenoscillations in Diffraction Theory* (John Wiley & Sons, Inc., New York 1999).

[54] R. Bennett, T.M. Barlow, and A. Beige, A physically motivated quantization of the electromagnetic field, Eur. J. Phys. **37** 014001 (2016).

[55] G. W. Hanson, Aspects of quantum electrodynamics compared to the classical case: Similarity and disparity of quantum and classical electromagnetics, IEEE Antennas and Propagation Magazine, **62,** 16, (2020).
19